# The mixed-valent manganese [3×3] grid [Mn(III)$_4$Mn(II)$_5$(2poap-2H)$_6$](ClO$_4$)$_{10}$·10H$_2$O, a mesoscopic spin-1/2 cluster

*Oliver Waldmann,*[*,†] *Hans U. Güdel,*[†] *Timothy L. Kelly,*[‡] *Laurence K. Thompson*[‡]

Department of Chemistry and Biochemistry, University of Bern, CH-3012 Bern, Switzerland,

Department of Chemistry, Memorial University, St. John's, NF, Canada A1B 3X7.

*To whom correspondence should be addressed. E-mail address: waldmann@iac.unibe.ch.

The magnetic susceptibility, and low-temperature magnetization curve, of the [3×3] grid [Mn(III)$_4$Mn(II)$_5$(2poap-2H)$_6$](ClO$_4$)$_{10}$·10H$_2$O (**1**) is analyzed within a spin Hamiltonian approach. The Hilbert space is huge (4,860,000 states), but the consequent use of all symmetries and a two-step fitting procedure nevertheless allows the best-fit determination of the magnetic exchange parameters in this system from complete quantum mechanical calculations. The cluster exhibits a total spin S = 1/2 ground state; the implications are discussed.

**Introduction**

The magnetism of polynuclear complexes containing magnetic metal ions, often called molecular nanomagnets, has captured the imagination of chemists and physicists alike. In the chemical arena, the building of novel materials with a functionality of potential interest for applications from a "bottom up" approach has stimulated much effort. From the physical perspective, these nanometer-sized magnetic clusters have been demonstrated to exhibit many spectacular magnetic quantum phenomena.[1-4]

The understanding of the magnetic properties of clusters with multiple magnetic centers, which usually starts with an analysis of the temperature dependence of the magnetic susceptibility $\chi(T)$, faces many challenges in order to evaluate the exchange coupling constants. On the one hand, the structure of the complex at hand easily may suggest many exchange parameters in the microscopic spin Hamiltonian,[5] and frequently, even with the use of simplified models, this leads to a heavily over-parameterized situation concerning the magnetic susceptibility. The only solution to this problem is to obtain information from several complementary experimental techniques. On the other hand, the dimension of the Hilbert space of the microscopic spin Hamiltonian increases exponentially with the spin dimension of the magnetic centers, so that the (numerical) calculation of magnetic properties quickly reaches the limits of today's computers. This is particularly true for the calculation of $\chi(T)$, since at higher temperatures essentially all energy levels are thermally populated and hence contribute, so that the full energy spectrum needs to be calculated. This is in contrast to other techniques for determining the magnetic parameters, which typically are performed at low temperatures, e.g., inelastic neutron scattering. Here, only a small number of the low-lying states is involved which with sparse-matrix diagonalization techniques can be obtained for systems orders of magnitude larger than those accessible by full diagonalization techniques.

In this work we analyze the magnetic susceptibility, and low-temperature magnetization curve, of the mixed-valent manganese [3×3] grid [Mn(III)$_4$Mn(II)$_5$(2poap-2H)$_6$](ClO$_4$)$_{10}$·10H$_2$O (**1**).[6,7] The structure is shown in Figure 1. The four spin-2 Mn(III) ions are located at the corners of the grid, while the remaining five metal sites are occupied by spin-5/2 Mn(II) ions. The Hilbert space of this cluster with its almost 5 million states is discouragingly large. We will show, however, that the subsequent use of all symmetries of the appropriate exchange Hamiltonian in combination with a two-step fitting procedure allows a reliable estimation of the exchange coupling parameters in this system (for the assumed exchange model see Figure 1b).

The experimental and theoretical analysis demonstrates that the antiferromagnetic interactions in **1** result in a total spin S = 1/2 cluster ground state. This can be understood within a simple classical picture of the ground-state spin configuration, in which the spin vectors on the Mn(III) sites and the central Mn(II) ion point up and the ones on the Mn(II) edge sites point down, accommodating the antiferromagnetic interactions best. Hence, (4×2) + (5/2) – (4×5/2) = 1/2. Thus, the grid **1** is a rare example of a mesoscopic spin-1/2 cluster, in which the S = 1/2 ground state arises from the concerted motion of many (magnetic) electrons. The prototypical example is the cluster $V_{15}$, in which 15 electrons couple to a (two-fold degenerate) S = 1/2 ground state.[8] Another example is the $Cr_7Ni$ molecule, in which 23 electrons act together to yield a S =1/2 ground state.[9,10] In **1**, the S = 1/2 ground state is the result of 41 electrons interacting in a completely isotropic fashion within the antiferromagnetic grid structure. The interest in this special class of magnetic molecules comes from recent theoretical work, which suggests that such objects might be suitable for building qubits, the elementary building blocks in quantum computers (in this context they have been denoted as "antiferromagnetic cluster qubits").[11,12]

**Experimental Section**

$[Mn(III)_4Mn(II)_5(2poap-2H)_6](ClO_4)_{10}·10H_2O$ (**1**) was synthesized as reported.[6,7] The magnetic moment of powdered and polycrystalline samples was measured with a MPMS5 SQUID magnetometer (Quantum Design). The polycrystalline samples were produced by taking crystals out of the mother liquor, putting them directly into grease, in which they were crushed. This procedure minimizes potential problems due to drying of the sample through solvent loss, thus yielding the most reliable magnetic data. However the weight of the samples cannot be determined reliably, and the data were calibrated by matching the susceptibility at high temperatures to that of known powder samples. The accuracy of absolute values for the molar susceptibility and magnetic moments were estimated to be about 5%. Preliminary magnetic data were communicated previously.[6]

**Results and Analysis**

The temperature dependence of the susceptibility, as determined from a measurement at a field of 0.5 T, and the magnetization curve at 2 K are shown in Figure 2. The maximum in $\chi(T)$ at about 60 K clearly indicates antiferromagnetic interactions in the cluster, and the strong increase at the lowest temperatures indicates a ground state with S > 0. The $\chi T$ value at 250 K is 27.3 cm$^3$ K mol$^{-1}$, which is significantly lower than that of five spin-5/2 and four spin-2 ions (33.89 cm$^3$ K mol$^{-1}$). This further demonstrates the antiferromagnetic interactions. At low temperatures, $\chi T$ approaches a value of 0.42 cm$^3$ mol$^{-1}$ indicative of a S = 1/2 ground state (0.375 cm$^3$ K mol$^{-1}$). The magnetization curve at 2 K further supports a S = 1/2 ground state. The continuing rise of the magnetization at higher fields suggests the presence of excited levels at about 10 K above the ground state.

On the basis of the grid structure, the magnetism of **1** should be well approximated by the microscopic spin Hamiltonian

$$H = -J_R \left( \sum_{i=1}^{7} \mathbf{S}_i \cdot \mathbf{S}_{i+1} + \mathbf{S}_8 \cdot \mathbf{S}_1 \right) - J_C \left( \mathbf{S}_2 + \mathbf{S}_4 + \mathbf{S}_6 + \mathbf{S}_8 \right) \cdot \mathbf{S}_9 \qquad (1)$$

where $J_R$ describes the next-neighbor exchange interactions between the Mn(II) and Mn(III) ions on the grid periphery, and $J_C$ the interactions between the edge Mn(II) ions and the central Mn(II) ion. The S = 1/2 ground state implies that both $J_R$ and $J_C$ are antiferromagnetic, i.e., $J_R$, $J_C$ <0. $S_2 = S_4 = S_6 = S_8 = S_9 = $ 5/2 and $S_1 = S_3 = S_5 = S_7 = 2$. This corresponds to a [3×3] grid of five Mn(II) ions and four Mn(III) ions, with the Mn(III) ions located at the corners of the grid consistent with the structure. The dimension of the Hilbert space of this system is as large as 4,860,000. An exact (numerical) calculation of the energy spectrum, as required for the calculation and interpretation of the magnetic susceptibility data, is thus challenging, and one has to take advantage of the symmetries of the microscopic spin Hamiltonian as much as possible. The spin rotational symmetry of Hamiltonian (1) allows one to work with a spin level basis set, where each level is classified by the quantum numbers of the total spin, S and M. For Hamiltonian (1), the Hilbert space consists of a total of 398,400 spin levels, and the largest matrix to be diagonalized has a dimension of 49,995 (see Table 1). This still by far exceeds the capabilities of modern personal computers (a memory exceeding 23 GB would be required). However, the [3×3] grid structure exhibits an idealized $D_4$ spatial symmetry, which manifests itself as a $D_4$ spin permutational symmetry of Hamiltonian (1).[13] Accordingly, the basis functions can be chosen to also transform according to the irreducible representations $A_1$, $A_2$, $B_1$, $B_2$, and each of the components of E of the group $D_4$. A numerical efficient implementation of the spin permutational symmetry, however, is possible only for a coupling scheme of the spins which is left invariant under the operations of the group elements of $D_4$.[13] In the present case, this requirement is fulfilled, e.g., for $\mathbf{S}_{15} = \mathbf{S}_1 + \mathbf{S}_5$, $\mathbf{S}_{37} = \mathbf{S}_3 + \mathbf{S}_7$, $\mathbf{S}_{1357} = \mathbf{S}_{15} + \mathbf{S}_{37}$, $\mathbf{S}_{26} = \mathbf{S}_2 + \mathbf{S}_6$, $\mathbf{S}_{48} = \mathbf{S}_4 + \mathbf{S}_8$, $\mathbf{S}_{2468} = \mathbf{S}_{26} + \mathbf{S}_{48}$, $\mathbf{S}_{12345678} = \mathbf{S}_{1357} + \mathbf{S}_{2468}$, and $\mathbf{S} = \mathbf{S}_9 + \mathbf{S}_{12345678}$. The resulting factorization of the Hilbert space is given in Table 1 (further details of the factorization procedure are given in the supporting information). The dimension of the largest matrix is now reduced to 12,486, which is still rather large but can be well handled on present day personal computers with 2 GB memory storage. A single calculation for one set of the parameters $J_R$ and $J_C$ requires about 2 days on a fast modern personal computer with 2 GB of RAM.

A full least-squares fitting of the magnetic susceptibility data, in which both $J_R$ and $J_C$ are allowed to vary independently, is thus unrealistic. However, it is possible within a reasonable time frame to fit the susceptibility data with Hamiltonian (1) for a fixed ratio of $J_C/J_R$: The Hamiltonian is rewritten as $H = -J_R \left( H_R + J_C/J_R H_C \right)$, with obvious meanings of $H_R$ and $H_C$, and the energy spectrum is calculated for $J_R = 1$ and a given ratio $J_C/J_R$. The energy spectrum for any value of $J_R$ is then obtained by simply scaling the calculated energy values by $J_R$. The susceptibility is then easily determined via the Van-Vleck equation (see eq. 2, second-order terms do not appear here since an isotropic model is considered). Thus, a best-fit value for $J_R$ can be obtained with standard least-squares fitting routines, once the energy spectrum for a fixed ratio of $J_C/J_R$ has been calculated.

In order to obtain best-fit values for both $J_R$ and $J_C$, a two-step procedure was followed. In a first step, the energy spectrum was calculated for a number of values for $J_C/J_R$ (specifically 0.1, 0.2, 0.3, 0.4, 0.5, 0.6, 0.8, 1.0, 3.0, and 10), and the susceptibility data least-square fitted to the model

$$\chi(T) = \frac{N_A \mu_B^2 g^2}{3 k_B T} \frac{\sum_{S\alpha}(2S+1)S(S+1)\exp\left[-J_R E_{S\alpha}^0/(k_B T)\right]}{\sum_{S\alpha}(2S+1)\exp\left[-J_R E_{S\alpha}^0/(k_B T)\right]} + \chi_0. \qquad (2)$$

Here, the three parameters $J_R$, g and $\chi_0$ were allowed to vary independently. In eq 2, the sum runs over all spin levels, numbered by S and α, and $E_{S\alpha}^0$ refers to the energies of the spin levels for a given value of $J_C/J_R$. The constant $\chi_0$ accounts for a small diamagnetic background due to the grease in the sample. Plotting the goodness-of-fit parameter chi$^2$ as function of $J_C/J_R$ then reveals a best-fit value for $J_C/J_R$. This way one obtains the best-fit values for $J_R$ and $J_C$ independently. In order to estimate their confidence limits, it is necessary, in a second step, to again least-square fit the data for each ratio $J_C/J_R$, but now with the parameters g and $\chi_0$ set to their best-fit values (g = 2.11, $\chi_0$ = -0.006 cm$^3$ mol$^{-1}$, it is remarked that the absolute value of the g factor is of little significance in view of the 5% accuracy of the data calibration).[14]

The goodness-of-fit parameter chi$^2$ as function of $J_C/J_R$ is shown in Figure 3a. Chi$^2$ does not exhibit a simple parabolic dependence, as expected for a Gaussian statistical analysis, but instead shows a more trough-like behavior with a "bottom" reaching from $J_C/J_R \approx 0.3$ to about 0.8. The standard procedure of calculating estimated standard deviations (esds) is related to the curvature of the parabola approximating the chi$^2$ behavior near the minimum.[14] Since the curvature at the bottom is very small, the resulting calculated esds are ridiculously large, and thus do not provide reliable estimators for the confidence limits. In order to give an impression of what the chi$^2$ values refer to, the corresponding susceptibility curves are drawn for some of the of $J_C/J_R$ ratios in Figure 3b. For chi$^2$ values outside the trough, the susceptibility curves clearly deviate from the experimental data, but for $J_C/J_R$ values within the trough, the curves are statistically indistinguishable. Accordingly, $J_C/J_R$ = 0.55(10) is estimated, and with the best-fit value of $J_R$ = -12 K for this ratio, this finally translates into

$$J_R = -12(1) \text{ K}, J_C = -6.5(10) \text{ K}. \tag{3}$$

The susceptibility curve corresponding to these values (and the g, $\chi_0$ values given above) reproduces the experimental data very well (see Figure 2a).

For an isotropic Hamiltonian, such as Hamiltonian (1), the magnetic moments for arbitrary temperature and magnetic fields can also be calculated exactly from the zero-field energy spectrum,[13]

$$m(T,B) = \mu_B g \frac{\sum_{S\alpha} S B_S(gSx)\sinh[g(S+\tfrac{1}{2})x]\exp[-E_{S\alpha}/(k_B T)]}{\sum_{S\alpha} \sinh[g(S+\tfrac{1}{2})x]\exp[-E_{S\alpha}/(k_B T)]} \tag{4}$$

where $B_S(y)$ is the Brillouin function, $x = \mu_B B/(k_B T)$, and $E_{S\alpha}$ denotes the energy levels (in this work $E_{S\alpha} = J_R E^0_{S\alpha}$). It is easily confirmed that for B $\rightarrow$ 0 eq 4 reduces to the Van Vleck equation for the susceptibility. Equation 4 enables the calculation of the magnetization curve, and the result is in excellent agreement with experiment, Figure 2b. The calculated energy spectrum yields a S = 1/2 ground state of the cluster, in agreement with the data, and a S = 3/2 level at 9.5 K, which explains the upturn of the m(B) curve at higher fields. Thus, as a conclusion, the obtained best-fit values reproduce the magnetism of **1** very well. With a Hilbert space of dimension 4,860,000, this is to date by far the largest system for which a full quantum mechanical analysis of the magnetization curves could be achieved.

On general grounds it cannot be assumed that magnetic anisotropy terms in the microscopic spin Hamiltonian, for instance due to ligand-field or dipole-dipole interactions, are negligible.[15,3] However, in the present case their effects are hardly seen in measurements on powder (or polycrystalline) samples. On the one hand, effects of the anisotropy are detectable only at the lowest temperatures, since at higher temperatures, as soon as an appreciable number of levels become thermally populated, the anisotropy effects average out. On the other hand, the S = 1/2 ground state of **1** cannot exhibit a zero-field-splitting, i.e., the anisotropy is not effective in the ground state. This in our opinion explains why the simple isotropic Hamiltonian (1) manages to reproduce well both the susceptibility and the magnetization curve.

**Discussion**

It is interesting to inspect the calculated energy spectrum for **1**. The full spectrum, as a function of S, is shown in Figure 4a, and a more detailed view of the low-energy part is provided in Figure 4b. As mentioned already, the ground state belongs to S = 1/2, followed by a S = 3/2 state at 9.5 K. The higher-lying levels show a remarkable, but well-known structure,[16,17,2,4] where the lowest states for each S exhibit a quadratic increase of energy $\propto S(S+1)$, which is characteristic for rotational bands. The lowest rotational band of states is known as the L band or "tower of states". Furthermore, the next-higher lying states above the L band form another set of rotational bands, also showing the typical S(S+1) increase in energy. This set of rotational bands has been collectively denoted as the E band.

A comparison shows, that the spectrum of **1** looks very similar to that of the antiferromagnetic wheels or the "original" Mn(II)-[3×3] grid.[17,2,4] In fact, its low-lying energy levels exhibit all the characteristic features found in these systems. The main difference is that in **1** the L band starts from S = 1/2, while it starts from S = 0 in the case of the wheels and S = 5/2 in the case of Mn-[3×3]. The similarity has the important implication that the spin dynamics, or elementary magnetic excitations, respectively, in **1** are explained by the same physical picture as in the wheels and Mn(II)-[3x3].[2,4] In this picture, the L band corresponds to the (quantized) rotation of the Néel vector, and the E band to the (quantized) spin-wave excitations. It has become clear in recent years, that this structure of the low-energy part of the spectrum is intimately connected to a "classical" spin structure.[17,2,4] Indeed, the S = 1/2 ground state can be easily rationalized by the classical spin configuration shown in Figure 1c.

These considerations implicitly demonstrate that the spin dynamics in **1** at the lowest temperatures are well described in terms of a Néel vector (the Néel vector is simply a vector which is parallel to the magnetization of one of the antiferromagnetic sublattices, e.g., to the up-pointing spins in Figure 1c). Thus, if the magnetic anisotropy is of the easy-axis type and large enough, and the Mn(III) ions are known to be good sources for easy-axis magnetic anisotropy, **1** would be in the regime of quantum tunneling of the Néel vector.[18,19] A careful determination of the magnetic anisotropy of **1** will be thus of high interest. As an additional comment, the classical spin structure also ensures that the effective (three-sublattice) spin Hamiltonian

$$H_{AB9} = -J'_R \mathbf{S}_A \cdot \mathbf{S}_B - J'_C \mathbf{S}_B \cdot \mathbf{S}_9 \qquad (5)$$

developed for the Mn(II)-[3×3] grid works well also for **1**.[19] In eq 5, A and B refer to the two magnetic sublattices formed by the edge and corner spins, i.e., $\mathbf{S}_A = \mathbf{S}_{1357}$, $\mathbf{S}_B = \mathbf{S}_{2468}$ (where $\mathbf{S}_A$, $\mathbf{S}_B$ assume their maximal values $S_A = S_1 + S_3 + S_5 + S_7$, $S_B = S_2 + S_4 + S_6 + S_8$; $S_A = 8$, $S_B = 10$ for **1**). Physically, this means that the low-energy dynamics corresponds to a motion in which the spins on each of the sublattices A and B act as a single, larger spin.

It is also interesting to look at **1** from another perspective, which is suggested by its $S = 1/2$ ground state. A spin-1/2 is a natural candidate for a quantum bit (qubit), the basic element of a quantum computer. However, among the many obstacles to be overcome in the realization of a quantum computer, one is the problem of addressing the qubit, which is a prerequisite for its initialization and read-out. For conventional spin-1/2 systems, in which the spin-1/2 arises from one unpaired electron, addressing is extremely difficult due to the typical smallness of the objects. However, recently it has been argued that mesoscopic spin-1/2 systems, in which the spin-1/2 arises from the concerted action of many electrons (41 in the case of **1**), might be good candidates for producing qubits (then called cluster qubits), because their larger physical size simplifies the task of addressing accordingly.[11]

A molecule discussed much in this context is the Cr$_7$Ni wheel.[9,10,12] In this wheel, the eight metal centers are arranged as an almost perfect ring and exhibit next-neighbor antiferromagnetic interactions.

Because of the smaller spin of the Ni(II) ion (spin-1) as compared to the Cr(III) centers (spin-3/2), the ground-state spin configuration is not fully compensated, resulting in a S = 1/2 ground state. The next-higher lying state, a S = 3/2 level, is at about 13 K above the ground state. Detailed numerical calculations and theoretical considerations have shown that the S = 1/2 cluster ground state of $Cr_7Ni$ indeed may provide a qubit, i.e., that the leakage to the nearby S = 3/2 levels is small enough, etc.[12]

The above discussion has shown that the classical spin structure in **1** also means that the effective 3-sublattice spin Hamiltonian concept describes the low-lying excitations well. Furthermore, in Ref. 19 it has been demonstrated, that for $J_C \gtrsim 0.01 J_R$ the sublattice spin $S_B$ and the central spin $S_9$ are so strongly coupled, that they act as combined spin. As a result, the 3-sublattice spin Hamiltonian can be further reduced to another effective spin Hamiltonian, which is exactly the effective spin Hamiltonian of a modified antiferromagnetic wheel. Thus, magnetically, **1** behaves at low temperatures exactly like a modified wheel with a S = 1/2 ground state, i.e., like $Cr_7Ni$. Also the energy gaps to the next-higher lying S = 3/2 are on the same order (9.5 K in **1** and 13 K in $Cr_7Ni$). The considerations drawn for $Cr_7Ni$ in the context of the applicability as cluster qubits[12] thus are valid also for **1**. In short, the mixed-valent manganese [3×3] grid **1** might be another system with significant potential as a cluster qubit. Recently, it has been shown that suitably functionalized manganese [3×3] grids can be organized in monolayers of surface-bound molecules onto substrates, e.g. Au(III), and can be individually addressed by scanning probe techniques,[6] overcoming another prerequisite for their application.

**Acknowledgments**

Partial financial support from the EC-RTN-QUEMOLNA, contract n° MRTN-CT-2003-504880, the Natural Sciences and Engineering Research Council of Canada, and the Swiss National Science Foundation is gratefully acknowledged.

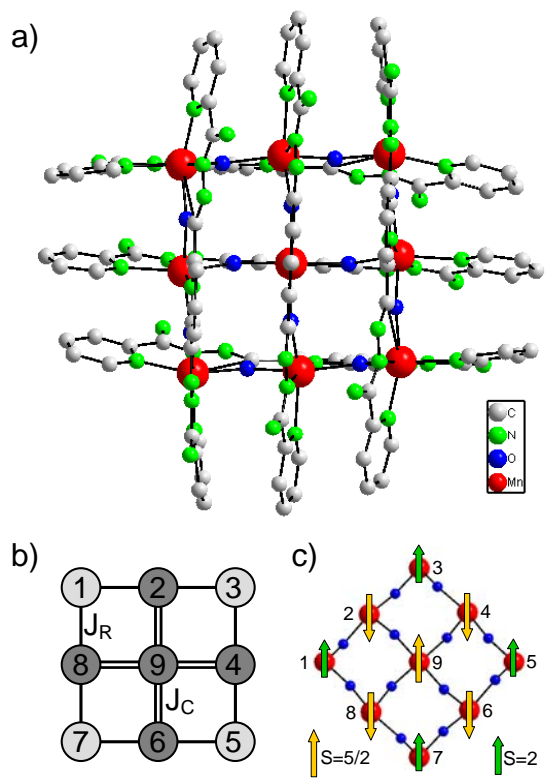

**Figure 1.** (a) Structural representation of the cation in **1**. (b) Magnetic exchange model for **1**. The light-gray circles represent the spin-2 Mn(III) ions, the dark-gray circles the spin-5/2 Mn(II) ions. (c) Classical spin configuration of the S = 1/2 ground state in **1**.

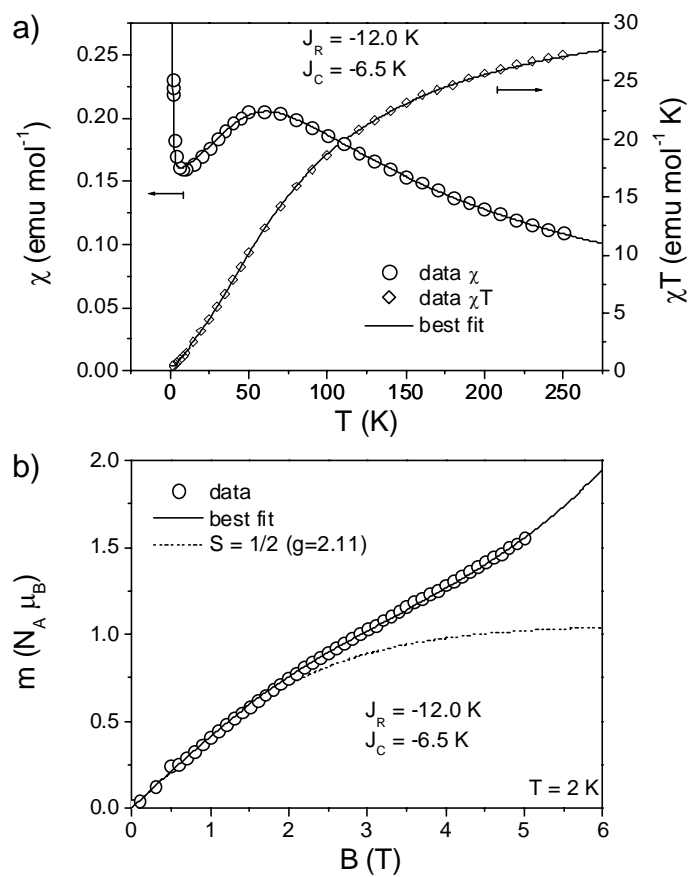

**Figure 2.** The open symbols in (a) show the magnetic susceptibility vs. temperature and in (b) the magnetic moment vs. applied field at 2 K of **1**. The solid line represents the best-fit results as calculated from Hamiltonian (1) with the exchange parameters as indicated, the dashed line in (b) the magnetization curve for a S = 1/2 level.

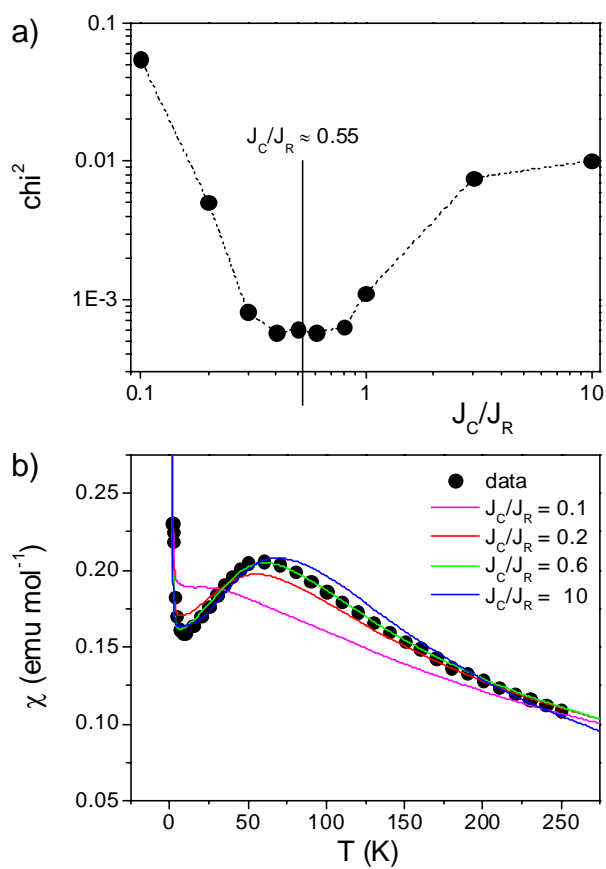

**Figure 3.** (a) Goodness-of-fit parameter chi$^2$ as function of the ratio $J_R/J_C$ (for details see text). (b) Comparison of the experimental $\chi(T)$ curve (circles) with the best-fit results for the indicated values of $J_R/J_C$ (solid lines).

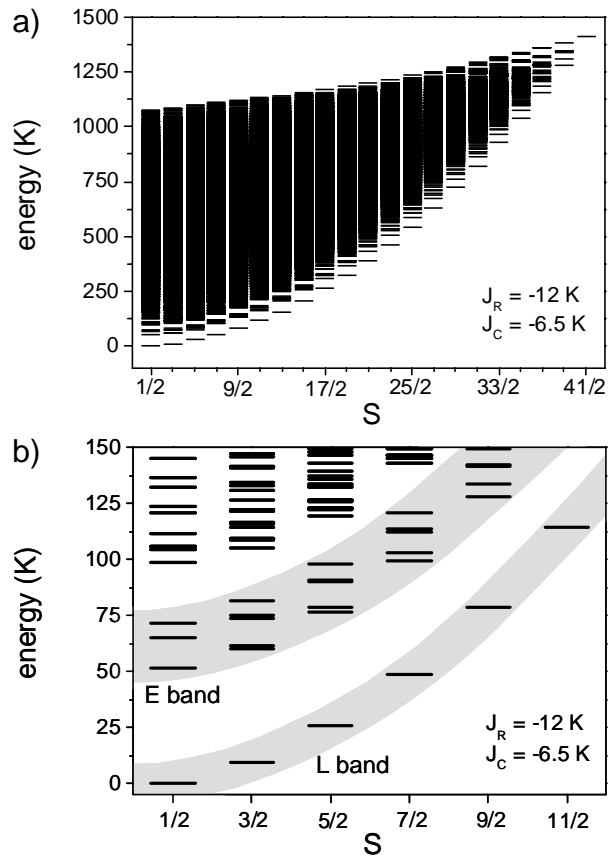

**Figure 4.** Energy spectrum vs. total spin quantum number S as calculated from Hamiltonian (1) with the parameters indicated in the panels. (a) Full energy spectrum. (b) Detailed view on the low-energy sector, highlighting the L band ($\hat{=}$ quantized rotation of the Néel vector) and E band ($\hat{=}$ quantized spin wave excitations).

**Table 1.** Classification scheme for the mixed-valent manganese [3×3] grid **1** in the D$_4$ symmetry group.

| S | A$_1$ | A$_2$ | B$_1$ | B$_2$ | E | total |
|---|---|---|---|---|---|---|
| 1/2 | 2032 | 1990 | 2011 | 2006 | 4013 | 16065 |
| 3/2 | 3828 | 3747 | 3794 | 3781 | 7575 | 30300 |
| 5/2 | 5212 | 5095 | 5162 | 5149 | 10291 | 41200 |
| 7/2 | 6052 | 5908 | 5991 | 5969 | 11960 | 47840 |
| 9/2 | 6340 | 6174 | 6263 | 6246 | 12486 | 49995 |
| 11/2 | 6100 | 5925 | 6025 | 6000 | 12025 | 48100 |
| 13/2 | 5482 | 5302 | 5402 | 5385 | 10757 | 43085 |
| 15/2 | 4603 | 4432 | 4529 | 4506 | 9035 | 36140 |
| 17/2 | 3647 | 3485 | 3571 | 3557 | 7103 | 28466 |
| 19/2 | 2704 | 2563 | 2642 | 2625 | 5267 | 21068 |
| 21/2 | 1897 | 1772 | 1840 | 1831 | 3648 | 14636 |
| 23/2 | 1240 | 1140 | 1195 | 1185 | 2380 | 9520 |
| 25/2 | 768 | 685 | 727 | 723 | 1436 | 5775 |
| 27/2 | 436 | 376 | 408 | 404 | 812 | 3248 |
| 29/2 | 235 | 189 | 213 | 212 | 415 | 1679 |
| 31/2 | 113 | 84 | 99 | 98 | 197 | 788 |
| 33/2 | 53 | 33 | 42 | 42 | 80 | 330 |
| 35/2 | 20 | 10 | 15 | 15 | 30 | 120 |
| 37/2 | 8 | 2 | 5 | 5 | 8 | 36 |
| 39/2 | 2 | 0 | 1 | 1 | 2 | 8 |
| 41/2 | 1 | 0 | 0 | 0 | 0 | 1 |

**Supporting Information**

Following the suggestion of a referee, we provide here more information, in a tutorial fashion, on the block factorization of the microscopic spin Hamiltonian (1) using the spin rotational and spin permutational symmetries (SRS and SPS). The general procedure has been described in detail in chapters II and IV of Ref. 13. Here it will be explored for the concrete example of the complex **1**. As in the main text, vectors are displayed as bold symbols, but operators, in contrast, will be explicitly identified by a hat.

In general, the procedure of block-factorizing the Hamiltonian matrix by employing the symmetry of the Hamiltonian consists of using symmetry-adapted wave functions as basis functions for setting up the Hamiltonian matrix. Group theory shows that blocks with different symmetry labels, i.e., blocks which transform according to different irreducible representations of the symmetry group of the Hamiltonian do not mix, leading to the block structure. Thus, the main task is to i) construct the appropriate symmetry-adapted wave functions and ii) to calculate the matrix elements of the Hamiltonian in this basis. In principle, the method for employing the SRS and SPS is exactly the same and based on the so-called "basis-function generating machine" (see below). The formalism for the SRS, however, is quite developed and known under the names of, e.g., the Racah formalism or the irreducible tensor operator (ITO) technique. These are well explained in many text-books and we assume here familiarity of the reader with these methods.

The SRS refers to the invariance of Hamiltonian (1) with respect to the total spin $\hat{\mathbf{S}} = \sum_i \hat{\mathbf{S}}_i$ (which implies the quantum numbers S and M) and is employed by using spin functions as basis functions. These are obtained by coupling the spins according to a spin coupling scheme such as $\hat{\mathbf{S}}_{15} = \hat{\mathbf{S}}_1 + \hat{\mathbf{S}}_5$, $\hat{\mathbf{S}}_{37} = \hat{\mathbf{S}}_3 + \hat{\mathbf{S}}_7$, $\hat{\mathbf{S}}_{1357} = \hat{\mathbf{S}}_{15} + \hat{\mathbf{S}}_{37}$, $\hat{\mathbf{S}}_{26} = \hat{\mathbf{S}}_2 + \hat{\mathbf{S}}_6$, $\hat{\mathbf{S}}_{48} = \hat{\mathbf{S}}_4 + \hat{\mathbf{S}}_8$, $\hat{\mathbf{S}}_{2468} = \hat{\mathbf{S}}_{26} + \hat{\mathbf{S}}_{48}$, $\hat{\mathbf{S}}_{1-8} = \hat{\mathbf{S}}_{1357} + \hat{\mathbf{S}}_{2468}$, $\hat{S} = \hat{S}_{1-8} + \hat{S}_9$, and are written as $|S_1S_5S_{15}S_3S_7S_{37}S_{1357}S_2S_6S_{26}S_4S_8S_{48}S_{2468}S_{1-8}S_9SM\rangle$ or in shorthand notation as $|\alpha SM\rangle$ (i.e., α stands for the string of intermediate quantum numbers).

The SPS refers to the spatial symmetry properties of Hamiltonian (1), or the point-group symmetry of the molecule under consideration. In order to employ the SPS via the basis-function generating machine one has to identify the symmetry operators, to calculate their action on the spin functions (wherewith automatically employing also SRS), and to know the representation matrices for the irreducible representations of the SPS group (knowledge of the character table is thus not sufficient if the group contains two and higher dimensional representations, but we assume here that the representation table is known, for the $D_4$ group it is given in Table S1).

Thus, one first has to become clear about the symmetry operations. It is clear that the point-group symmetry of the molecule shows up in some way as a further symmetry of Hamiltonian (1), additional to the SRS. There is, however, a subtlety because the elements of the point group act in 3-dimensional space, while the spin functions live in spin space. One way to deal with this is to use the point-group elements and to somehow "map" the spin functions to orbital space (this approach has been developed e.g. by Tsukerblat in a mathematically rigorous way). Here we follow the conceptually opposite method of "mapping" the point-group elements to the spin space, as outlined in Ref. 13. For most cases this approach is much easier to work with, though in some rare cases some subtle points may emerge (which however are irrelevant for isotropic spin models as considered here).

In this approach, the point-group symmetry of the molecule corresponds to a permutational symmetry of the microscopic spin Hamiltonian, i.e., the symmetry operations are permutations of the spin sites and the symmetry group is a subgroup of the permutation group (hence the notation SPS). For example for **1**, the $C_2$ symmetry axis perpendicular to the grid plane corresponds to the permutation 123456789 → 567812349, i.e., the spin site 1 becomes the new site 5, site 2 the new site 6, and so on. The permutational symmetry elements for the molecule **1** in the $D_4$ SPS group are displayed in Figure S1. As a side remark, molecules with different point-group symmetries thus may exhibit the same SPS. For instance, molecules with a $C_s$, $C_i$, or $C_2$ point group will all exhibit the same SPS group since these point-groups are isomorphic (they are different point groups since the operations $C_s$, $C_i$, or $C_2$ are different in 3-dimensional space, but all lead to the same permutation 12 → 21). In the present case, **1** exhibits a (approximate) $D_{2d}$ symmetry, which is isomorphic to $D_4$ (and $C_{4v}$). It is thus a matter of semantics to denote the SPS group of **1** as $D_4$ or $D_{2d}$.

Having identified the symmetry operations, one needs to know the effect of the corresponding symmetry operators $\hat{O}(P)$ on the spin functions $|\alpha SM\rangle$ (where P denotes one of the permutations of the SPS group). In general, the result will be a linear combination of several other spin functions, i.e., $\hat{O}(P)|\alpha SM\rangle = \sum_{\alpha'} c^S_{\alpha\alpha'}|\alpha'SM\rangle$. However, for cases where the spin coupling scheme is left invariant by the permutations P of the SPS group, the situation simplifies enormously as the application of $\hat{O}(P)$ on $|\alpha SM\rangle$ produces exactly one other spin function, multiplied by a phase factor:

$$\hat{O}(P)|\alpha SM\rangle = (-1)^{\kappa(P)}|\alpha'SM\rangle \tag{S1}$$

This is apparently an enormous simplification, and particularly convenient for numerical implementation. Fortunately, from the many possibilities to couple the individual spins to the total spin $\hat{S}$, in the case of **1** several spin coupling schemes fulfill the condition of invariance (as for instance the one given in the above; the invariance of the spin coupling scheme is demonstrated also in Figure S2 exemplarily for one permutation). The resulting string of intermediate quantum numbers $\alpha'$ is easy to infer, it is obtained by applying the permutation P on the string $\alpha$, i.e., $\alpha' = P\alpha$. For instance, for the permutation 123456789 → 567812349 (corresponding to the $C_2^z$ symmetry element of **1**) one finds $\alpha' = S_5 S_1 S_{15} S_7 S_3 S_{37} S_{1357} S_6 S_2 S_{26} S_8 S_4 S_{48} S_{2468} S_{1-8} S_9$. The determination of the exponent $\kappa(P)$ of the phase factor is straightforward but more involved; the procedure is described in Ref. 13, chapter IV.A.

Finally, the symmetry-adapted spin functions $|\alpha S M \Gamma \lambda\rangle$, where $\Gamma$ refers to one of the irreducible representations (irrep) of the SPS group and $\lambda$ to its components (in case the dimension of $\Gamma$ is larger than one), are constructed using the basis-function generating machine

$$|\alpha S M \Gamma \lambda\rangle = \frac{d_\Gamma}{h} \sum_P \Gamma^*_{\lambda\lambda}(P) \hat{O}(P) |\alpha S M\rangle, \tag{S2}$$

where $d_\Gamma$ is the dimension of the irrep $\Gamma$, h the number of elements in the SPS group (here h = 8), and $\Gamma_{\lambda\mu}(P)$ the matrix representation of the permutation P in the irrep $\Gamma$. With the symmetry-adapted spin functions as basis functions, the Hilbert space block-factorizes into blocks, which each can be labeled by S, M, $\Gamma$, $\lambda$, see Figure S3. That is, the Hilbert space decomposes into blocks for each S, which each in turn decompose into 2S+1 blocks corresponding to M = -S, …, S, which each further decomposes into 6 blocks, four for each 1-dimensional irreps and two for the 2-dimenional irrep E of $D_4$. Since all blocks with identical S and $\Gamma$ are identical by symmetry [where are hence $(2S+1)d_\Gamma$ such blocks] only one of them needs to be considered and diagonalized [if the obtained energies are counted $(2S+1)d_\Gamma$ times in order to get the full energy spectrum].

For the actual numerical diagonalization, the matrices in the various subspaces have to be set up, i.e., the matrix elements $\langle \alpha S M \Gamma \lambda | \hat{H} | \alpha' S M \Gamma \lambda \rangle$ calculated (resulting in the matrices $H^{SM\Gamma\lambda}_{\alpha\alpha'}$ for each tuple S, M, $\Gamma$, $\lambda$). Since for each S and $\Gamma$ all blocks with different M, $\lambda$ are equivalent, only one has to be calculated with M, $\lambda$ set to arbitrary values, e.g., M = S and $\lambda = 1$. Employing eq (S1) naively for the numerical calculation of $\langle \alpha S M \Gamma \lambda | \hat{H} | \alpha' S M \Gamma \lambda \rangle$ will produce very inefficient code since a double sum over all elements of the SPS group needs to be performed, so that in total $h^2$ matrix elements $\langle \alpha S M | \hat{H} | \alpha' S M \rangle$ would have to be calculated. However, employing the great orthogonality theorem of group theory the calculation can be reduced to a single sum,

$$\langle\alpha SM\Gamma\lambda|\hat{H}|\alpha'SM\Gamma\lambda\rangle = \frac{d_\Gamma}{h}\sum_P \Gamma_{\lambda\lambda}(P)\langle\alpha SM|\hat{H}\hat{O}(P)|\alpha'SM\rangle$$
$$= \frac{d_\Gamma}{h}\sum_P \Gamma_{\lambda\lambda}(P)(-1)^{\kappa(P)}\langle\alpha SM|\hat{H}|(P\alpha')SM\rangle \quad (S3)$$

which saves a factor h in computation time (which is a factor of 8 in the present case of **1**). Employing the SPS obviously permits to reduce the memory usage, but only with the "trick" eq (S3) also the computation time for the calculation of the whole energy spectrum can be reduced (otherwise the total computation time in fact would somewhat increase). The matrix elements $\langle\alpha SM|\hat{H}|\alpha''SM\rangle$ ($\alpha'' = P\alpha'$) involve only the "original" spin functions and are calculated via the standard ITO techniques.

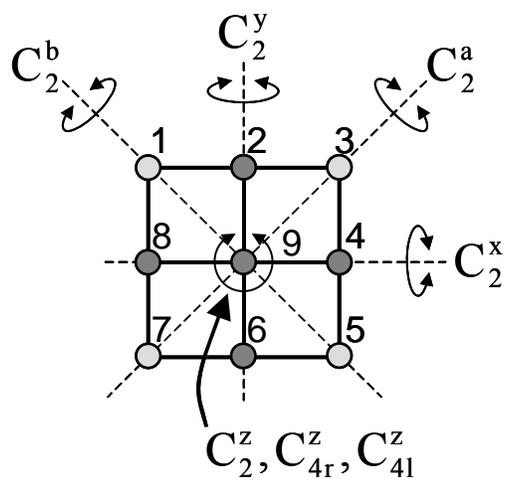

**Figure S1**: Symmetry elements of **1** in the $D_4$ spin permutational group.

$$123456789 \rightarrow 781234569$$

$\hat{\mathbf{S}}_{15} = \hat{\mathbf{S}}_1 + \hat{\mathbf{S}}_5$ $\quad\longrightarrow\quad$ $\hat{\mathbf{S}}_{37} = \hat{\mathbf{S}}_3 + \hat{\mathbf{S}}_7$

$\hat{\mathbf{S}}_{37} = \hat{\mathbf{S}}_3 + \hat{\mathbf{S}}_7$ $\quad\longrightarrow\quad$ $\hat{\mathbf{S}}_{15} = \hat{\mathbf{S}}_1 + \hat{\mathbf{S}}_5$

$\hat{\mathbf{S}}_{1357} = \hat{\mathbf{S}}_{15} + \hat{\mathbf{S}}_{37}$ $\quad\longrightarrow\quad$ $\hat{\mathbf{S}}_{1357} = \hat{\mathbf{S}}_{15} + \hat{\mathbf{S}}_{37}$

$\hat{\mathbf{S}}_{26} = \hat{\mathbf{S}}_2 + \hat{\mathbf{S}}_6$ $\quad\longrightarrow\quad$ $\hat{\mathbf{S}}_{48} = \hat{\mathbf{S}}_4 + \hat{\mathbf{S}}_8$

$\hat{\mathbf{S}}_{48} = \hat{\mathbf{S}}_4 + \hat{\mathbf{S}}_8$ $\quad\longrightarrow\quad$ $\hat{\mathbf{S}}_{26} = \hat{\mathbf{S}}_2 + \hat{\mathbf{S}}_6$

$\hat{\mathbf{S}}_{2468} = \hat{\mathbf{S}}_{26} + \hat{\mathbf{S}}_{48}$ $\quad\longrightarrow\quad$ $\hat{\mathbf{S}}_{2468} = \hat{\mathbf{S}}_{26} + \hat{\mathbf{S}}_{48}$

$\hat{\mathbf{S}}_{1-8} = \hat{\mathbf{S}}_{1357} + \hat{\mathbf{S}}_{2468}$ $\quad\longrightarrow\quad$ $\hat{\mathbf{S}}_{1-8} = \hat{\mathbf{S}}_{1357} + \hat{\mathbf{S}}_{2468}$

$\hat{\mathbf{S}} = \hat{\mathbf{S}}_{1-8} + \hat{\mathbf{S}}_9$ $\quad\longrightarrow\quad$ $\hat{\mathbf{S}} = \hat{\mathbf{S}}_{1-8} + \hat{\mathbf{S}}_9$

**Figure S2:** Demonstration of the invariance of the spin coupling scheme used for **1** (see text) for the example of the permutation $123456789 \rightarrow 781234569$ (= symmetry element $C_{4r}^z$). The equations on the left side represent the "original" spin-coupling scheme; the ones on the right side the coupling scheme obtained after applying the permutation (which consists of changing the indices of the spin operators accordingly). The spin coupling scheme is invariant if one gets back the same coupling rules, as is the case in the shown example.

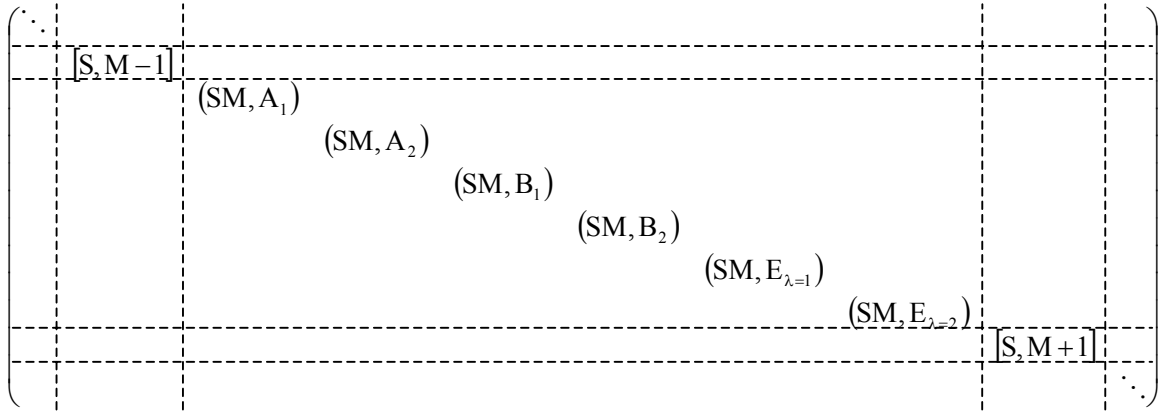

**Figure S3**: Block-factorization of the Hamiltonian matrix under the SRS and $D_4$ SPS. According to the SRS the Hamilton matrix decomposes into blocks for each S and M (for each S only one block with e.g. M = S needs to be diagonalized since they are identical for all M = -S, …, S). Each of these blocks further decompose into six sub blocks due to the $D_4$ SPS, four blocks for each of the 1-dimenisonal irreducible representations $A_1$, $A_2$, $B_1$, and $B_2$, and two blocks for the 2-dimensional irreducible representation E (of the two E blocks only one needs to be diagonalized since both are equivalent).

**Table S1.** Irreducible representations of the D$_4$ symmetry group. The permutations P associated to each symmetry element as appropriate for the molecule **1** are also given.

|       | E | $C_2^z$ | $C_{4r}^z$ | $C_{4l}^z$ | $C_2^x$ | $C_2^y$ | $C_2^a$ | $C_2^b$ |
|-------|---|---------|------------|------------|---------|---------|---------|---------|
| P     | 123456789 | 567812349 | 781234569 | 345678129 | 765432189 | 321876549 | 543218769 | 187654329 |
| A$_1$ | 1 | 1 | 1 | 1 | 1 | 1 | 1 | 1 |
| A$_2$ | 1 | 1 | 1 | 1 | -1 | -1 | -1 | -1 |
| B$_1$ | 1 | 1 | -1 | -1 | 1 | 1 | -1 | -1 |
| B$_2$ | 1 | 1 | -1 | -1 | -1 | -1 | 1 | 1 |
| E | $\begin{pmatrix} 1 & 0 \\ 0 & 1 \end{pmatrix}$ | $\begin{pmatrix} -1 & 0 \\ 0 & -1 \end{pmatrix}$ | $\begin{pmatrix} 0 & 1 \\ -1 & 0 \end{pmatrix}$ | $\begin{pmatrix} 0 & -1 \\ 1 & 0 \end{pmatrix}$ | $\begin{pmatrix} 0 & 1 \\ 1 & 0 \end{pmatrix}$ | $\begin{pmatrix} 0 & -1 \\ -1 & 0 \end{pmatrix}$ | $\begin{pmatrix} 1 & 0 \\ 0 & -1 \end{pmatrix}$ | $\begin{pmatrix} -1 & 0 \\ 0 & 1 \end{pmatrix}$ |